\documentclass[aps,prl,floatfix,twocolumn,nofootinbib,showpacs]{revtex4}
\usepackage{graphicx} 
\usepackage[sort&compress]{natbib}
\usepackage{graphicx}
\newcommand{\BEQ}{\begin{equation}}
\newcommand{\EEQ}{\end{equation}}
\newcommand{\BEA}{\begin{eqnarray}}
\newcommand{\EEA}{\end{eqnarray}}

\renewcommand{\d}{{\rm d}}

\newcommand{\Tr}{{\rm Tr}}

\newcommand{\CZ}{{\cal Z}}

\begin{document}
\draft
\title
{Statistical networks emerging from link-node interactions}

\author{
A.E. Allahverdyan and K.G. Petrosyan}
\address{Yerevan Physics Institute,
Alikhanian Brothers St. 2, Yerevan 375036, Armenia
}

\date{\today}

\begin{abstract} We study a model for a statistical network formed by
interactions between its nodes and links. Each node can be in one of two
states (Ising spin up or down) and the node-link interaction facilitates
linking between the like nodes. For high temperatures the influence of
the nodes on the links can be neglected, and we get the Ising
ferromagnet on the random (Erdos-Renyi) graph.  For low temperatures the
nodes get spontaneously ordered. Due to this, the connectivity of the
network enhances and links having a common node are correlated. The
emerged network is clustered.  The node-link interaction shifts the
percolation threshold of the random
graph to much smaller values, and the very percolation
transition can become of the first order: the giant cluster coexist with
the unconnected phase leading to bistability and hysteresis.  The model
can be applied to the striction phenomena in magnets and to studying
opinion formation in the sociophysical context.

\end{abstract}






\maketitle

Statistical mechanics of networks is a growing field with a wide
range of applications \cite{reviews}. The subject originated with
works of chemical physicists \cite{flory} and mathematicians
\cite{rapo}, while the modern trends focus on interdisciplinary
applications (biology, social sciences) \cite{reviews}.

Normally networks are modeled either via passive nodes with a given
distribution of links, or with a given dynamics of link formation
(growing network) \cite{reviews}. One of the basic models in the first
class is the random (Erdos and Renyi) graph \cite{reviews,rapo}: a set
of $N$ nodes with the links independently distributed between them.
The model adequately describes some (e.g., percolation) but not all the
relevant features of the real networks (e.g., clustering). This
motivated generalizations of the random graphs within statistical mechanics
of networks \cite{ext}.

Here we propose to model a network such that its nodes and links are active
variables influencing each other.  We present below possible applications of this approach.

Consider $N$ labeled nodes which carry Ising spins $\sigma=\{\sigma_i=\pm 1\}_{i=1}^N$.
Linking between these nodes is described by
the symmetric adjacency matrix $J=\{J_{ik}\}$:
$J_{ik}=J_{ki}=1(0)$ indicate on the presence (absence) of the corresponding
undirected link. No self-links are present $J_{ii}=0$.
The Hamiltonian of the system is
\BEA
H(\sigma,J)=
-\gamma{\sum}_{i<k} J_{ik}\sigma_i\sigma_k+\alpha{\sum}_{i<k} J_{ik},
\label{granada73}
\EEA
where $\gamma$ is the coupling of the link-node interaction, while the second term
with $\alpha>0$ generates (for $\gamma=0$) the proper random graph behavior.
For $\gamma>0$
(ferromagnet) linking between the like nodes
$\sigma_i\sigma_k>0$ is favored. Likewise,
if $J_{ik}=1$, then $\sigma_i$ and $\sigma_k$ tend to
line up.

We assume that {\it i)} $J$ and $\sigma$ are statistical systems
with different times-scales: spins (links) are fast
(slow); {\it ii)} $J$ and $\sigma$ may have different temperatures.  For
constructing the stationary distribution in the spirit of the
information-theoretical approach \cite{jaynes} we introduce the
entropies of the links and spins \cite{Onsager}: $S_J\equiv
-\Tr_JP(J)\ln P(J)$, $S_\sigma\equiv -\Tr_J[\,\Tr_\sigma P(\sigma|J)\ln
P(\sigma|J)\,]$, where $\Tr_J$ ($\Tr_\sigma$) is the summation over all
configurations of $J$ ($\sigma$).  $P(J)$ and $P(\sigma|J)$ are,
respectively, the probability of the links and the conditional
probability of the spins. Due to the above assumption ${\it i)}$
the relevant probability for the spins is
$P(\sigma|J)$, and thus $S_\sigma$ is the relevant {\it conditional
entropy}.  $P(J)$ and $P(\sigma|J)$ are found by minimizing
the average energy $U=\Tr_J\Tr_\sigma
H(\sigma,J)P(J)P(\sigma|J)$ for the fixed values of $S_J$ and
$S_\sigma$.  To this end minimize the Lagrange function
$U-T_\sigma S_\sigma-T_JS_J$, with
$T_\sigma=1/\beta_\sigma$ and $T_J=1/\beta_J$ being the
Lagrange factors or temperatures \cite{Onsager}:
\BEA
\label{imamali173}
P(\sigma |J)=\frac{e^{-\beta_\sigma H(\sigma,J)}}{Z(J)}, \quad
Z(J)\equiv\Tr_\sigma e^{-\beta_\sigma H(\sigma,J)},\\
P(J)=\CZ^{-1}\,{Z^{n}(J)},\quad
\CZ\equiv\Tr_J Z^{n}(J), \quad n\equiv T_\sigma/T_J,
\label{imamali273}
\EEA
where $-T_\sigma\ln Z(J)$ is an effective Hamiltonian
forming a Gibbs distribution at temperature $T_{J}$ for $P(J)$.

Alternatively, we can recover (\ref{imamali173}, \ref{imamali273}) via
the microscopic approach \cite{LandauerWoo,Coolen,Onsager}
subjecting $J$ and $\sigma$ to thermal baths at
temperatures $T_J$ and $T_\sigma$, respectively.

{\it Limiting cases.}
For $\gamma=0$ (no spins thus passive nodes) we get from $P(J)$
the standard random graph with the links $J_{ik}$ independently assuming values $0$ and $1$ with
the probabilities, respectively,
$1-\frac{c}{N}$ and
$\frac{c}{N}$. Here $c$ is finite for $N\gg 1$ and is found from
\BEA
\label{ku2}
\alpha\beta_J
\equiv\ln\left([N-c]/c\right).
\EEA
Since links can be formed between any pair of the nodes, most of
$J_{ik}$ have to be zero to ensure a finite average connectivity:
$\sum_{k}\langle J_{ik}\rangle=c$.

When $n=T_\sigma/T_J\to 0$, but $c$ (and thus $\alpha\beta_J$) is fixed,
the nodes do not react on the links which still form the random graph.
The Ising model on the random graph (which simulates disordered media)
is widely employed for studying magnets, lattice gases, etc,
\cite{BY,grest}. Finite $n$ describes situations, where the reaction of
the spins on the lattice is important, e.g., the striction effect. For ordered
lattices such effects were studied in detail \cite{larkin}; the present model
describes striction for a disordered lattice.

{\it Phase structure}. We calculate $\CZ$ in (\ref{imamali273})
via the notorious replica method \cite{BY}, i.e., first taking
$n=T_\sigma/T_J$ integer and then analytically continuing to a
real $n$:
\BEA
\CZ={\rm Tr}_{\sigma,J}
e^{-\beta\sum_{a=1}^nH(\sigma^{a}, J)}
={\rm Tr}_\sigma
e^{\frac{c}{N}
\sum_{i<k}
e^{\beta\gamma\sum_{a=1}^n\sigma^a_i\sigma_k^a}
}.\nonumber
\label{kaa}
\EEA
We now apply the identity ($\tau_a=\pm 1$)
\BEA
\label{durman}
&&e^{\beta\gamma \sum_{a=1}^n\tau_a}=b_0+\sum_{r=1}^n b_r
\sum_{a_1<...<a_r}
\tau_{a_1}...\tau_{a_r},\\
&&b_r={\rm Tr}_\tau\left(\tau_1...\tau_r
e^{\beta\gamma \sum_{a=1}^n\tau_a}\right)
=\tanh^r(\beta\gamma)\cosh^n(\beta\gamma),\nonumber
\EEA
and employ $e^{ax^2/2}=\int\frac{\d y}{\sqrt{2\pi}}
e^{-y^2/2+\sqrt{a}\, xy}$:
\BEA
\label{mish1}
&&\CZ=\int\prod_{a=1}^n\prod_{a_1<...<a_r}
\sqrt{Ncb_r}\,\d Q_{a_1...a_r}\, e^{\frac{cNb_0}{2}-NF(Q)},~~~~
\\
&&F(Q)\equiv
\frac{c}{2}{\sum}_{r=1}^n b_r
{\sum}_{a_1<...<a_r}
Q^2_{a_1...a_r}\nonumber\\
&&-\ln
\Tr_\sigma
\exp\left[
\sum_{r=1}^n
\sum_{a_1<...<a_r}
c\,b_r\,
Q_{a_1...a_r}\sigma_{a_1}...\sigma_{a_r}
\right].
\label{mish2}
\EEA
$F$ is the thermodynamic potential of the problem \cite{Onsager}.
The values of $Q_{a_1...a_r}$ are found from treating the integral in (\ref{mish1})
via the saddle-point method, where one searches the deepest minima of $F(Q)$
(necessarily, $\frac{\partial F}{\partial Q_{a_1...a_r} }=0$).
For situations of our interest the replica symmetry holds, i.e.,
$Q_{a_1...a_r}$ depend only on the number of the replicas:
$Q_{a_1...a_r}=Q^{(r)}$. Since all nodes are equivalent \cite{BY},
\BEA
Q_{a_1...a_r}=Q^{(r)}=\Tr_JP(J)\,\left[\Tr_\sigma P(\sigma|J
  )\,\sigma\right]^r.
\label{khazar}
\EEA
For an arbitrary $n$ we have to involve all
$Q_{a_1...a_r}$, which makes the explicit analysis rather difficult.
Things are simpler for $n=1,2$, where at best only two parameters
are involved $Q_1=M$ (magnetization) and $Q_{12}=Q$ (Edwards-Anderson parameter \cite{BY}).
For $n=1$
the assumption on the wide separation of characteristic times is irrelevant,
because $P(\sigma,J)$ is a Gibbsian at temperature $T_\sigma=T_J$ and does not depend
at all on the characteristic times.
We get from (\ref{mish1}, \ref{mish2}):
\BEA
\label{duba}
F= \frac{cb_1M^2}{2}-\ln[2 \cosh (cb_1M)],\,
~M=\tanh (cb_1M),
\EEA
where the behavior of $M$ is controlled by a single parameter $cb_1$:
for $cb_1<1$ the only solution is $M=0$, while for $cb_1\geq 1$ we have
a smooth (second-order) transition to the ferromagnetic phase, where
$M\not =0$. Eq.~(\ref{duba}) admits two ferromagnetic solutions with $\pm M$.
At the transition point one of these values is chosen randomly.
The choice can be made deterministic
by applying a small magnetic field $h$, i.e., by adding
$h\sum_{i=1}^N\sigma_i$ ($h\ll J$) to the Hamiltonian.
We can get $cb_1>1$ by increasing the spin-link coupling $\gamma$,
or decreasing $T_\sigma$
with constant $n$ and $c$ ({\it thermal} phase transitions).
The presence of
macroscopic ($\propto N$) cluster of connected nodes (giant component)
is necessary for a thermal phase transition.
In the {\it percolation limit} thermal fluctuations are absent:
$T_\sigma, T_J\to 0$, while $n$ and $cb_r$ are fixed. As seen from
(\ref{ku2}) this requires very small connectivity $c$.  Then $cb_r=c_e$
does not depend on $r$. Eq.(\ref{duba}) then shows that the ferromagnet appears via
second-order phase transition for $c_e=1$. The presence of the
giant component is necessary and sufficient for the percolational phase transition.

\begin{figure}
\includegraphics[width=0.79\linewidth]{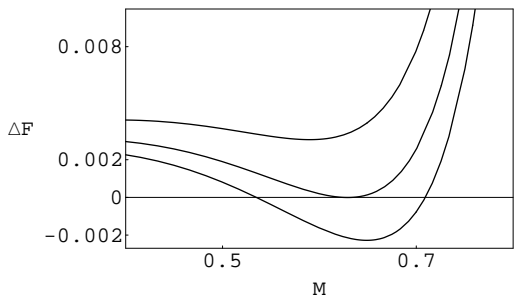}
\hfill
\caption{  The difference $\Delta F$ between the
ferromagnet and paramagnet values of $F$.
On each curve the minimum corresponds to the stationary
  magnetization $M$. The curves refer to different
  temperatures and $c=\gamma=1$; from the top to bottom: $T=1.452,\, 1.445,\,
  1.44$. The top-line minimum corresponds to metastable
  ferromagnet, the bottom-line one to the stable ferromagnet, while
  the middle line is drawn
  at the transition temperature $T=T_{2}$, where $\Delta F=0$.
}
\label{fig1}
\end{figure}

Second-order thermal transitions can be studied for an arbitrary $n$.
Next to such a transition $\Tr_\sigma [P(\sigma|J )\sigma_i]$
is small, and (\ref{khazar}) implies $1\gg M\gg Q$.
Expanding $F$ in $Q$ and $M$ (Ginzburg-Landau expansion) and
solving $\frac{\partial F}{\partial Q}=0$ so that
$Q=\frac{c^2}{1-cb_2}b_1b_2 M$ is expressed via $M$, gives
\BEA
\label{aqva}
F=A_0M^2+A_4M^4,~~~~~~~~~~~~~~~~~~~~~~~~~~~~~~~~~~~~~~\\
A_0=\frac{cb_1n}{2}(1-cb_1),\,~
A_4=\frac{nc^4b_1^4[1-cb_2(3n-2)]}{12(1-cb_2)}.
\EEA
At high temperatures both $A_0$ and
$A_4$ are positive, and $F$ is minimal for $M=Q=0$ (paramagnet).
$A_0$ changes its sign at $cb_1=1$, i.e, at
$ct=(1-t^2)^{\frac{n}{2}}$, where $t\equiv\tanh \gamma\beta_\sigma$.
Then $A_4>0$ is equivalent to $1>(3n-2)t$.  Thus for $n<1$, $A_4>0$ for
$A_0\simeq 0$ that means a second-order transition to ferromagnet:
at $cb_1\geq 1$, $F$ in (\ref{aqva}) changes from monostable to the
bistable shape with non-zero $M$ and $Q$.  For $n>1$ this behavior is
true for sufficiently large $c$, e.g., for $n=2$ conditions $ct=(1-t^2)$ and
$1>(3n-2)t$ lead to $c>15/4$.  For smaller $c$, however, $A_4$ changes
its sign before $A_0$ which indicate on first-order transition as the
next case $n=2$ shows. Here we have three order parameters
$Q_1$, $Q_2$ and $Q_{12}=Q$. The deepest
minima of $F$ corresponds to
$Q_1=Q_2=M$,
\BEA
\label{gabor}
F=cb_1M^2+\frac{cb_2Q(Q-2)}{2}
-\ln(2\varphi),~~~~~~~~~~~~~~~~~~~\\
\varphi M={\sinh(2cb_1M)},~
\varphi Q={\cosh(2cb_1M)-e^{-2cb_2Q}},\,
\label{k}
\EEA
where Eqs.(\ref{k})
is derived from $\frac{\partial F}{\partial M}=\frac{\partial F}{\partial Q}=0$,
and where $\varphi\equiv e^{-2cb_2Q}+\cosh(2cb_1M)$.
For $Q\not =0$ and $M\not =0$, Eqs.(\ref{k}) are solved to give
($\theta\equiv \tanh (2c b_1M)$)
\BEA
\label{adel}
Q=\frac{1}{2cb_2}\ln\left[
\frac{M\sqrt{1-\theta^2}}{\theta-M}
\right],\quad
M=\frac{\theta (Q +1)}{2},
\EEA
Eq.(\ref{adel}.1) is put into (\ref{gabor}) getting $F$ as a function of
$M$. Plotting this function for various $\gamma$ (or $T_\sigma$, where
$c$ and $n=2$ are fixed) we select those minima of $F$ which
satisfy (\ref{adel}.2); see Fig. \ref{fig1}.  At some temperature
$T_\sigma=T_{1}$ the ferromagnet
with $Q>0$ and $M>0$ first appears as a metastable phase (local
minimum of $F$). For $T<T_{2}$ this state provides a value of $F$ lower
than the paramagnetic $-\ln 2$. $T_2$ is thus the first-order
phase-transition temperature, since $Q$ and $M$ jump at this point from
$Q=M=0$ to some finite values.
During this transition paramagnet remains
locally stable; it, however, gets unstable for $cb_1>1$, i.e., at some
lower temperature $T_{3}<T_{2}$, where $T_{3}$ is determined from
$cb_1=1$.  As an example: for $c=1$, $T_{1}=1.457$, $T_{2}=1.445$, and
$T_{3}=1.385$. For $c\to 15/4$ all these three temperatures merge to
$T_{3}$.
In contrast to the thermal transitions, for the considered case $n=2$
the percolation transition (driven by $cb_1=cb_2=c_e$) is always
first-order: at $c_{e}^{(1)}=0.805$ there first
appears the metastable ferromagnet with $Q>0$ and $M>0$.  For
$c_e>c_{e}^{(2)}=0.824$ it is truly (globally) stable. The paramagnet
is (locally) stable till $c_e=1$.

{\it The structure of the network} in the present model
is a collective effect, since spins react on the links.
The average of $m$ distinct links is derived analogously to (\ref{mish1}):
\BEA
\langle {\prod}_{\alpha=1}^m J_{i_{\alpha}k_{\alpha}}
\rangle=\frac{c^m}{N^m}\langle e^{\beta\gamma\sum_{\alpha=1}^m\sum_{a=1}^n
  \sigma^a_{i_{\alpha}}
  \sigma^a_{k_{\alpha}}}\rangle.
\label{hh}
\EEA
The transformation of (\ref{hh}) is
illustrated for $m=1$:
{\it i)} Apply (\ref{durman}); {\it ii)} Recall that
the model does not have any space structure and that all the links are equivalent,
i.e, the linking probability $\langle J_{ik}\rangle=\Delta_1$ can be sought for as
$\frac{1}{N^2}\sum_{i\not =k}\langle J_{ik}\rangle$. {\it iii)}
Neglect fluctuations of the macroscopic
quantities, e.g.,
$\langle
(\sum_{j=1}^n\sigma_j)^2
\rangle=
\langle
\sum_{j=1}^n\sigma_j
\rangle^2$.
This general point of the statistical mechanics
can be proven directly for the present model. The final formula
reads
\BEA
\Delta_1=\frac{c}{N}\left[
b_0+nb_1M^2+\frac{n(n-1)Q^2}{2}
\right],\quad n=1,2.
\EEA
In the paramagnet $\Delta_1=cb_0/N$ corresponds to the usual random
graph with the effective connectivity $c_e=cb_0$. It increases|by jump
for the first-order transition|in the ferromagnet. The reason for this increase
is that once two spins are (in average) lined up, they
get linked stronger. The same tendency is
seen for the probability of two links. Instead of this quantity, it is
more convenient to study the two link correlator $\langle
J_{ik}J_{lj}\rangle- \langle J_{ik}\rangle\langle J_{lj}\rangle$. It
is zero if $J_{ik}$ and $J_{lj}$ do not have a common node. This holds
as well for all higher order correlators of $J$'s and is explained as
follows: from (\ref{imamali173}, \ref{imamali273},
\ref{granada73}) we see that the
conditional probability for $J$ factorizes, $P(J|\sigma)
=\prod_{i<k}P(J_{ik}|\sigma)$, because there is no direct $J-J$
interaction in the Hamiltonian (\ref{granada73}).
Thus $J$'s can be correlated only due to fluctuations of the spins from one
realization to another.

The two-link correlator $\delta_2=\langle J_{ik}J_{jk}\rangle- \langle
J_{ik}\rangle\langle J_{jk}\rangle$ ($i\not =j$) is conveniently
discussed for $n=1$ (the more general and tedious expression leads to
similar conclusions):
\BEA
\delta_2=\frac{c^2}{N^2}\,b_1^2(1-M^2)M^2\geq 0.
\EEA
$\delta_2$ is zero in the paramagnet, and thus $\langle
J_{ik}J_{jk}\rangle=c^2b_0^2/N^2$ is again of the random graph form.
In the ferromagnet $\delta_2$ is {\it positive}: if the nodes
$i$ and $k$ are linked stronger than in average, $i$ strongly links
with $j$. $\delta_2>0$ also means that
the links are attracted to the common node, since
$\Delta_1\delta_2= {\rm Pr}[J_{ik}=1|J_{jk}=1]-{\rm Pr}[J_{ik}=1]$.

The probability $\Delta_3\equiv\langle J_{ij}J_{jk}J_{ki}\rangle$
to have three links forming a triangle is representative for
$n=2$:
\BEA
\Delta_3
=\frac{c^3}{N^3}\,\{\,b_0^3+2b_1^3+b_2^3
+6QM^2b_1^2\left[
b_0+
b_2+
2b_1
\right]+
\nonumber\\
6M^2b_1\left[ b_1(b_0+ b_2)+ b_2^2+ b_0^2 \right] +3Q^2\left[
2b_1^3+b_0b_2^2+ b_2b_0^2 \right]\}. \nonumber \label{hhh3} \EEA In
the paramagnet $\Delta_3 =\frac{c^3}{N^3}\,\{\,b_0^3+2b_1^3+b_2^3\}$
clearly deviates from the random graph behavior ${c^3b_0^3}/{N^3}$.
In the ferromagnet $\Delta_3$ increases, by a jump if the transition
is of the first order. The effect of clustering (informally: friends
of our friends are our friends) is when two nodes already connected
via a third node tend to establish a direct connection. It is
characterized through a coefficient $K={\rm
Pr}[J_{ij}=1|J_{ik}=1,J_{jk}=1]-{\rm Pr}[J_{ij}=1]$, which is zero
for the random graph. For our case
$K\equiv\frac{c^3}{N^3}\,\{2b_1^3+b_2^3\}>0$ in the paramagnet, thus
defining a type of short-range order. The random graph behavior is
recovered only for very high temperatures where
$b_r=\cosh^n(\beta\gamma)\tanh^r\beta\gamma\to 0$ for $r>0$. In the
ferromagnet $K$ first increases, but goes to zero for $Q,\,M\to 1$.
The case $n=1$|where the system is in global equilibrium and the
time-scale separation is irrelevant|is exceptional, since here the
clustering in the paramagnet is absent:
$K=b_1^2(b_1+2b_0)M^2(1-M^2)$.

Another relevant quantity for the network is the degree
$d_i\equiv\sum_kJ_{ik}$ of node $i$
(note $\langle d_i\rangle=\Delta_1$). For the random graph the degrees
of different nodes correlate very weakly, $\langle d_id_k\rangle-\langle
d_i\rangle\langle d_k\rangle\propto 1/N$, and the distribution
of $d_i$ is Poissonian: $P(d)=e^{-c}\frac{c^d}{d!}$ \cite{reviews}. The
first feature holds in our model due to domination in $\langle
d_id_k\rangle$ of links without a common node. The second feature holds
only in the paramagnet; see \cite{ak} for details.

{\it Discussion.} We studied a model for statistical network, where both
nodes and links are active variables influencing each other. The nodes
can be in two states (spin up or down) and the like nodes tend to link.
Important aspects of the model are {\it i)} time-scale separation: the
nodes change faster than the links and {\it ii)} non-equilibrium: the
temperatures of the nodes ($T_\sigma$) and links ($T_J$) are different.
The phase structure of the model crucially depends on whether the
reaction of the nodes on the links is more ($n\equiv T_\sigma/T_J>1$) or less
($n\leq 1$) pronounced.  For the first case the spins
modify the links and create persistency: the
phase transition to the ferromagnet is
of first order; paramagnet (ferromagnet) is metastable below (above) the
transition point. This bistability implies hysteresis and memory: when
changing the temperature not very slowly, the final state of the system
(para or ferro) depends on its initial state. The transition
becomes second-order, realized by instability of the paramagnet, either if
the network is dense, or if $n\leq 1$.  For $n\to 0$
no reaction of the nodes on the links is present, and we return to Ising
ferromagnet on the (Erdos-Renyi) random graph \cite{grest}.

The network is clustered already in the paramagnet: two nodes
connected via a third one, tend to link directly. In the ferromagnet
there are link-link correlations: two links having a common node
attract each other. The connectivity (linking probability),
clustering and link-link correlations increase during the phase
transition and can serve as alternative order parameters.

The model can simulate the magnetostriction effect for disordered
magnets. For the Ising model on regular lattices the effect was
studied in Refs.\cite{larkin} and used to explain experiments on
crystals of ${\rm MnAs}$, ${\rm NH_4Cl}$, etc, which also display
striction-driven change of the ferromagnetic transition order from
the second to the first.

Yet another application can be modeling the opinion formation, a
sociological subject that recently got an input of physical ideas
\cite{op}.  Here the spin $\sigma_i=\pm 1$ refers to two opinions
each agent may have. It is influenced by the (social) noise at temperature
$T$.  The concrete manifestation of ``agents'' and ``opinions'' may
be voting, propagation of fashion, investment, etc. The agents
interact within slowly changing social network $J$. The
ferromagnetic interaction describes the herding-collaboration:
like-minded agents tend to connect; linked agents impose their
opinion on one another. The linking costs energy, as given by the
term $\propto \alpha$ in (\ref{granada73}). As compared to various
applications of statistical physics models to the opinion formation
\cite{op}, our approach emphasizes the collaboration aspect of the
like-minded agents and the dynamical character of the involved
social network; see \cite{dit} for somewhat related ideas.
Here is an outline of our results in the opinion formation
context.

The paramagnetic phase corresponds to a plural society, where each
agent has his own opinion.  Herding and collaboration are not
efficient either because linking is too costly, or noises are too
strong. Still, there is a short-range order: three (not less!)
agents tend to connect and to share one opinion.  For weaker noises
or less costly linking the herding behavior starts to dominate. If
the influence of the agents on the linking is not pronounced ($n<1$)
the majority of agents coherently forms a single opinion driven by
weak external influences (spontaneous second-order transition). Below the
transition, almost no agent disagrees with the majority. For
pronounced influence on the linking ($n>1$) the plural society
survives the first-order transition (i.e., jump-like dominance of a
single opinion) and slowly decays on a long time-scale. It gets unstable upon further
decrease of the linking cost.  We see that revolutionary
(jump-like) changes can occur in response to smooth variations of
external conditions in societies with strong trends to establish
groups of like-minded agents.  Such changes can be prevented if the
infrastructure network is dense enough.  Within the single-opinion
dominated society strongly linked agents tend to be linked even
stronger, and tend to link directly with indirectly related agents.
The connectivity in this phase is larger than in the plural phase.

A.E. A. was supported by CRDF grant ARP2-2647-YE-05.
K.G. P. was supported by ISTC grant A-820.


\begin{thebibliography}{99}


\bibitem{reviews} S.N. Dorogovtsev and J.F.F. Mendes, Adv. Phys., {\bf
51}, 1079, (2002).  R. Albert and A.L. Barabasi, Rev. Mod. Phys., {\bf
74}, 47 (2002).  M.E.J. Newman, SIAM Review, {\bf 45}, 167, (2003).

\bibitem{flory}P.J. Flory, {\it Principles of Polymer Chemistry},
(Cornell Univeristy, Ithaca, 1953; chapter 9).

\bibitem{rapo}B. Bollob\'as, {\it Random Graphs}, (Academic Press, New York, 1985).

\bibitem{ext}J. Berg and M. Lassig, Phys. Rev. Lett., {\bf 89}, 228701  (2002).
J. Park and M.E.J. Newman, Phys. Rev. E, {\bf 70}, 066117  (2004).
M.E.J. Newman, {\it et al.},\,
Phys. Rev. E 64, 026118 (2001).
M.E.J. Newman, ArXiv: cond-mat/0202208.
Z. Burda, {\it et al.}, ArXiv: cond-mat/0312494.

\bibitem{jaynes}E. Jaynes, Phys. Rev., {\bf 106}, 620 (1957).

\bibitem{Onsager} A.E. Allahverdyan and Th.M. Nieuwenhuizen,
Phys. Rev. E, {\bf 62}, 845 (2000).

\bibitem{LandauerWoo} R. Landauer and J. Woo, Phys. Rev. A {\bf 6}, 2205 (1972).

\bibitem{Coolen} A.C.C. Coolen, {\it et al.}, 
J. Phys. A {\bf 26}, 3681 (1993).
A.E. Allahverdyan, {\it et al.}, 
Eur. Phys. J. B, {\bf 16}, 317, (2000).

\bibitem{BY}K. Binder and A.P. Young, Rev. Mod. Phys., {\bf 58}, 801 (1986).

\bibitem{grest}
M.J. Stephen and G.S. Grest, Phys. Rev. Lett., {\bf 38}, 567  (1977).
L. Viana and A.J. Bray, J. Phys. C {\bf 18}, 3037 (1985).


\bibitem{larkin} C. Bean and D. Rodbell, Phys. Rev., {\bf 126}, 104
(1962).  A.I. Larkin and S.A. Pikin, Sov. Phys. JETP {\bf 29}, 891
(1969).  C.P. Slichter, {\it et al.}, Phys. Rev. B, {\bf 4}, 907 (1971).

\bibitem{ak}A.E. Allahverdyan and K.G. Petrosyan, in preparation.

\bibitem{op}
K. Sznajd-Weron and J. Sznajd, Int. J. Mod. Phys. C, {\bf 11}, 1157 (2000).
P. L. Krapivsky and S. Redner, Phys. Rev. Lett. 90,
238701 (2003).
Q. Michard and J.-P. Bouchaud,
arXiv:cond-mat/0504079.
Curty and M. Marsili,
arXiv:physics/0506151.

\bibitem{dit}
D. Stauffer,{\it et al.}, arXiv:cond-mat/0402670.

\end{thebibliography}
\end{document}